# Laser Profile Changes Due to Photon-Axion Induced Beam Splitting

By: Carol Scarlett
Florid A&M University

**ABSTRACT:**
This paper looks at a potentially unique measurable due to photon-axion coupling in an external magnetic field. Traditionally, detection of such a coupling has focused on observation of an optical rotation of the beam's polarization due to either a birefringence or a path length difference (p.l.d.) between two polarization states. Such experiments, utilizing mirror cavities, have been significantly limited in sensitivity; approaching coupling strengths of $\sim g_a = 10^{-7}$ GeV$^{-1}$. Here the bifurcation of a beam in a cavity is explored along with the possibility of measuring its influence on the photon density. Simulations indicate that coupling to levels $g_a \sim 10^{-12}$ are, with an appropriate choice of cavity, within measurable limits. This is due to a rapid growth of a signal defined by the energy loss from the center accompanying an increase in the region beyond the beam waist. Finally, the influence of a non-zero axion mass is explored.

## I.  Introduction:

The search for axions has lead to recent theoretical development suggesting the production of a pseudo state, photon-axion mixing, in the presence of an external B field gradient[1,3]. Here a possible method of detecting these coupled states is discussed. This work builds heavily on the theoretical framework presented in reference [3].

Traditionally, experimental searches for axions have tried to detect optical rotations of one component of the beam's polarization. Such an effect could be generated by either a difference in velocity or path length for the two directions of polarization. Cavity experiments, see references [4,7], have focused rotations. Their sensitivities have been limited to refractive index differences of $\sim 10^{-20}$ corresponding to photon-axion couplings of $\sim 10^{-7}$ GeV$^{-1}$.

In addition to rotating the photon beam's polarization, a coupling between axions and photons can lead to beam splitting. This theory was originally applied to astrophysical objects such as magnetars[2] and used to calculate a measurable splitting of microwave radiation, in the presence of such strong fields, visible as a delayed pulse. Here, a terrestrial experiment is simulated and the measurable is defined as a change in density profile of the beam, visible on a ccd camera.

## II.  Approximating Beam Shifting

As an alternative to searching for changes in the polarization due to photon-axion coupling, consider how the formation of a pseudo state that splits the energy of an incoming beam into two new beams changes the number of photons in a region defined by the beam's waist. In theory[3], the two outgoing beams separate in an external inhomogeneous field, acquiring equal and opposite transverse momentum. The density of photons in the waist is diminished proportional to the splitting angle and the length of the external field. A signal can be defined by the difference between the integrated density of the beam waist for the case of maximum, applied external field and for the case of zero external field. Finally, modulation of the external field enhances the experimental sensitivity by giving a narrow frequency range in which to look for a signal.

To understand this analytically, start with a Gaussian function that describes the distribution of photons $P_D$ in the laser beam:

$$P_D = Ae^{-\frac{1}{2}[\frac{x}{r}]^2} \qquad (1)$$

Where x represents the position from the center of the distribution, implicitly zero in the expression above, r defines the beam waist, and $2 \cdot r$ represents the $e^{-2}$ position where the beam is ~ 1/7 of its maximum. If both splitting and broadening occur due to coupling of the beam to an axion field, the expression above can be rewritten in terms of these changes:

$$P_D' = A \cdot \frac{r}{r+\varepsilon} \cdot \frac{1}{2} \cdot e^{-\frac{1}{2}[\frac{x-\alpha}{r+\varepsilon}]^2}$$

$$P_D'' = A \cdot \frac{r}{r+\varepsilon} \cdot \frac{1}{2} \cdot e^{-\frac{1}{2}[\frac{x+\alpha}{r+\varepsilon}]^2} \qquad (2)$$

Where $\alpha$ represents a displacement of the beam's center in a positive direction from zero and $\varepsilon$ represents an increase in the beam's divergence due to the photon-axion coupling and/or the additional path traveled by the split density. The last expression can be rewritten by expanding:

$$P_D' = A \cdot \frac{r}{r+\varepsilon} \cdot \frac{1}{2} \cdot e^{-\frac{1}{2}[\frac{x}{r} \cdot (1-\frac{\alpha}{x}) \cdot (1+\frac{\varepsilon}{r})^{-1}]^2}$$

$$P_D'' = A \cdot \frac{r}{r+\varepsilon} \cdot \frac{1}{2} \cdot e^{-\frac{1}{2}[\frac{x}{r} \cdot (1+\frac{\alpha}{x}) \cdot (1+\frac{\varepsilon}{r})^{-1}]^2} \qquad (3)$$

Applying a Taylor expansion yields:

$$(1+\frac{\varepsilon}{r})^{-1} \approx 1 - \frac{\varepsilon}{r} + \frac{\varepsilon^2}{r^2} + Higher-Order-Terms$$

$$P_D' \approx A \cdot (1-\frac{\varepsilon}{r}) \cdot \frac{1}{2} \cdot e^{-\frac{1}{2}[\frac{x}{r} \cdot (1-\frac{\alpha}{x}) \cdot (1-\frac{\varepsilon}{r}+...)]^2} \approx A \cdot \frac{r-\varepsilon}{2 \cdot r} \cdot e^{-[\frac{x^2}{r^2} \cdot (\frac{1}{2}-\frac{\alpha}{x}+\frac{\alpha^2}{x^2}-\frac{\varepsilon}{r}+...)]}$$

$$P_D'' \approx A \cdot (1-\frac{\varepsilon}{r}) \cdot \frac{1}{2} \cdot e^{-\frac{1}{2}[\frac{x}{r} \cdot (1+\frac{\alpha}{x}) \cdot (1-\frac{\varepsilon}{r}+...)]^2} \approx A \cdot \frac{r-\varepsilon}{2 \cdot r} \cdot e^{-[\frac{x^2}{r^2} \cdot (\frac{1}{2}+\frac{\alpha}{x}+\frac{\alpha^2}{x^2}-\frac{\varepsilon}{r}+...)]} \qquad (4)$$

For the case being considered here, both the splitting and the broadening will scale as the coupling constant $g_a$ (~ $10^{-12}$) thus the expression above can be written as:

$$P_D' \approx A \cdot \frac{r-\varepsilon}{2 \cdot r} \cdot e^{-\frac{1}{2} \cdot \frac{x^2}{r^2}} \cdot e^{+\frac{x^2}{r^2} \cdot \frac{\varepsilon}{r}} \cdot e^{-\frac{\alpha^2}{r^2}} \cdot e^{+\frac{x^2}{r^2} \cdot \frac{\alpha}{x}}$$

$$P_D'' \approx A \cdot \frac{r-\varepsilon}{2 \cdot r} \cdot e^{-\frac{1}{2} \cdot \frac{x^2}{r^2}} \cdot e^{+\frac{x^2}{r^2} \cdot \frac{\varepsilon}{r}} \cdot e^{-\frac{\alpha^2}{r^2}} \cdot e^{-\frac{x^2}{r^2} \cdot \frac{\alpha}{x}}$$

(5)

The last expressions can be combined and subtracted from the initial distribution to reveal the degree of redistribution of the photon density.

$$P_D = A e^{-\frac{1}{2} \cdot \frac{x^2}{r^2}}$$

$$P_D'' + P_D' \approx A \cdot \frac{r-\varepsilon}{r} e^{-\frac{1}{2} \cdot \frac{x^2}{r^2}} \cdot e^{+\frac{x^2}{r^2} \cdot \frac{\varepsilon}{r}} \cdot e^{-\frac{\alpha^2}{r^2}} \cdot \cosh(\frac{x^2}{r^2} \cdot \frac{\alpha}{x})$$

(6)

$$P_D - (P_D' + P_D'') \approx A e^{-\frac{1}{2} \cdot \frac{x^2}{r^2}} [1 - \frac{r-\varepsilon}{r} \cdot e^{+\frac{x^2}{r^2} \cdot \frac{\varepsilon}{r}} \cdot e^{-\frac{\alpha^2}{r^2}} \cdot \cosh(\frac{x^2}{r^2} \cdot \frac{\alpha}{x})]$$

Finally, one derives an expression directly proportional to both the splitting angle (which is proportional to the coupling constant $g_a$) and the broadening due to increased divergence (also proportional to $g_a$). The final form rises exponentially with $\varepsilon$ and as a cosh function with an envelope that changes as the variable $\alpha$.

Conventionally, if one uses a Fabry-Perot cavity where a beam propagates back and forth over essentially the same path, one can use matrix formulism to track the changes in the beam profile. Here, however, <u>if shifting of the beam's center occurs then there are two vectors that must be tracked through the cavity: first, the vector defining the complex curvature (waist) of the beam and, second, a vector defining the position and angle for the central rays (momentum direction for the beam as a whole)</u>. Below, broadening is neglected, thus only the second vector is needed to calculate a measurable signal.

In the absence of broadening, equation 6 can be rewritten:

$$P_D - (P_D' + P_D'') \approx A e^{-\frac{x^2}{r^2}} [1 - (1 - \frac{\alpha^2}{r^2}) \cdot \cosh(\frac{x^2}{r^2} \cdot \frac{2 \cdot \alpha}{x})]$$

(7)

where the change in the density rises as a function of the shifting of each new beam from the original beam's center ($\alpha$). A first order expansion of the cosh shows that this rise is $\sim \alpha^2$. Here $\alpha$ can be expressed in terms of the splitting angle, cavity length, d, and a function $f(n)$ that measures the effective shifting of the new distributions away from the center:

$$\alpha = \theta_{split} \cdot d \cdot f(n) \tag{8}$$

The function $f(n)$ is extracted from simulations of the photon density. Ultimately, this function characterizes how a signal builds in a cavity and accounts for real effects such as focusing.

To track the effects of periodic focusing, a simulation using a confocal cavity was performed. Starting well within the sensitivity range of previous axion searches[6-7], $g_a \sim 10^{-6}$ GeV$^{-1}$, the parameters below are used:

Table 1:

| Cavity Type | Cavity Length | Magnetic Field Length | VB Strength | Laser Wavelength | Laser Energy | Mirror Radius | Number of Bounces | $\theta_{split} \sim 10^{-10}$ ($g_a = 10^{-6}$) |
|---|---|---|---|---|---|---|---|---|
| Confocal | 14 m | 10 m | 200 T/m | 1064 nm | 1 W | 25 m | $1.2 \cdot 10^4$ | $4 \cdot 10^{-10}$ |

The cavity here is a Fabry-Perot. The magnetic field is centered in the cavity and extends 10 m. To reduce background, the field is modulated and, to avoid interaction between the changing field and mirrors, there is a 2 m gap between either end of the magnet and the mirrors.

With the set of parameters in Table 1, it is possible to estimate the change in density, for a single traversal through the field, the $f(n) = 1$ case, over $\pm r$, the beam's radius. Evaluating equation (7), the density at a given value of position (x), an estimate from x = 0 to the x = 0.75 mm yields:

$$P_D - (P_D' + P_D'') \approx A e^{-\frac{x^2}{r^2}} [1 - (1 - \frac{\alpha^2}{r^2}) \cosh(\frac{x^2}{r^2} \cdot \frac{2 \cdot \alpha}{x})] \approx -A e^{-\frac{x^2}{r^2}} [\frac{\alpha^2}{r^2} - \frac{2 \cdot x^2}{r^2} \cdot \frac{\alpha^2}{r^2}]$$

At $x = 0$ the change is $\frac{\alpha^2}{r^2}$, while at $x \approx r$ the change is 0.

Approximating as a triangle: base = $\frac{1}{2} \cdot \frac{\alpha^2}{r^2}$ and height $\approx \frac{5}{3} \cdot 10^{+18}$ : (9)

$$P_D - (P_D' + P_D'') \approx \frac{5}{6} \cdot 10^{+18} [\frac{4 \cdot 10^{-10} \cdot 14}{7.5 \cdot 10^{-4}}]^2 \approx 9.3 \cdot 10^7$$

Where A has been normalized over the interval.

This difference can be compared to the shot noise level of $\sim 1.29 \cdot 10^9$. One should keep in mind this calculation is for a single traversal of the field with an integrated time of a single second.

### III. Simulating Splitting

The ABCD formalism was used to track the position of each new distribution created through photon-axion interactions. NOTE: short hand methods, such as Pascal Triangles, can predict the angular distribution of the light, however, interpreting the entries as the positions of emerging beams results in a non-conservation of transverse momentum (see Appendix). The Pascal

Triangle fails to predict the position needed to calculate a signal. Thus, in the simulations presented here, the vector defining the center of the beam and its direction (referred to as the chief ray) is carefully tracked through each traversal of the cavity.

The chief ray is tracked geometrically. The ABCD formalism to describe transit through various optical elements can be written as:

$$\text{Focusing:} \quad \begin{bmatrix} 1 & 0 \\ -\frac{1}{f} & 1 \end{bmatrix} \quad (10)$$

where f represents the focal length of a lens.

$$\text{Propagation:} \quad \begin{bmatrix} 1 & d \\ 0 & 1 \end{bmatrix} \quad (11)$$

Here d is the distance of propagation. To account for splitting a new matrix is introduced. When applied, this matrix takes one vector and transforms it into two vectors emerging from a single point, headed in distinct directions:

$$\text{Splitting:} \quad \begin{bmatrix} 1 & 0 \\ \pm \frac{\theta_{split}}{X_0} & 1 \end{bmatrix} \quad (12)$$

Where $X_0$ represent the current position before the matrix is applied and $\theta_{split}$ represents the splitting angle. Due to the curved path that the beam must take through space, the photons act as if they have an effective charge and experience a force in the presence of an external, inhomogeneous magnetic field[3], an angular adjustment is made:

$$\text{Angular Enhancement:} \quad \begin{bmatrix} 1 & 0 \\ (\pm \frac{\theta_{split}}{X_0}) & 1 \end{bmatrix} \quad (13)$$

Upon exiting the field, the chief rays propagate along a line tangent to this curve or with an angle twice that of the splitting within the field. Thus the last matrix is necessary to double the angle made by the chief ray as it emerges from the field.

Care must be taken when applying the last two matrices. These matrices do not get applied to the complex vector defining the beam waist and curvature as this would imply a broadening effect that is not the same effect as splitting. Unlike changes in the divergence, splitting imparts positive transverse momentum to a fraction of the light while imparting negative transverse momentum to the same fraction regardless of where the rays are relative to the beam center.

**Case 1: A Confocal Cavity**

Consider the geometry of the cavity displayed in Figure 1 below:

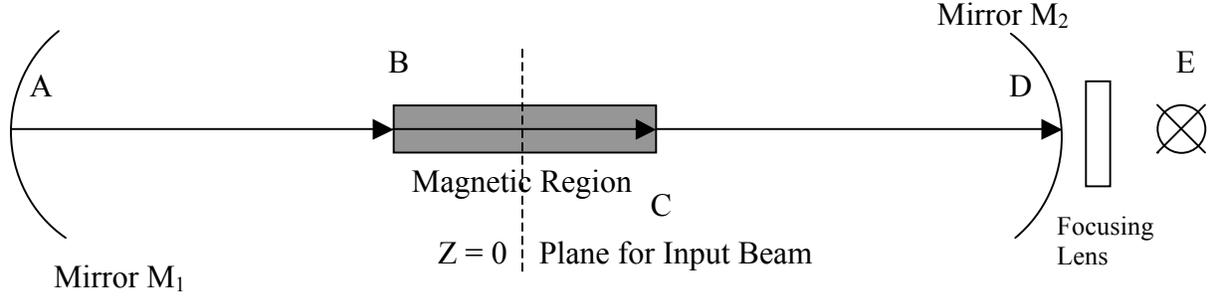

Figure 1: Sketch for a possible experimental setup showing a mirror cavity (A to D), magnetic field region (B to C) and detector (E).

A single traversal of the field involves a propagation from the first mirror $M_1$ labeled as position A, to the start of the magnetic field region at point B, to the end of this region at point C, to the second mirror $M_2$ labeled as position D and, finally, to a detector at point E – a distance of 2 m from the exit mirror. For the case of multiple traversals, the beam is focused at $M_2$ then travels through the cavity back to position A before either exiting or being reflected. The exiting distribution is focused onto the detector by a lens placed 0.5 m upstream of the mirror cavity.

In a longer note, where broadening is considered, transformation of the complex curvature is accounted for. For the case presented here, the plane where the beam waist is minimized, Z = 0, occurs at the center of the cavity. The matrix transformation defining a single traversal becomes:

$$\begin{bmatrix} 1 & (E-D) \\ 0 & 1 \end{bmatrix} \begin{bmatrix} 1 & (D-C) \\ 0 & 1 \end{bmatrix} \begin{bmatrix} 1 & 0 \\ \frac{(\pm)\theta_{split}}{R^*} & 1 \end{bmatrix} \begin{bmatrix} 1 & (C-B) \\ 0 & 1 \end{bmatrix} \begin{bmatrix} 1 & 0 \\ \frac{\pm\theta_{split}}{R^*} & 1 \end{bmatrix} \begin{bmatrix} 1 & (B-A) \\ 0 & 1 \end{bmatrix} \begin{bmatrix} R_0 \\ \theta_0 \end{bmatrix} \quad (14)$$

Where $R_0$ is the initial position and $\theta_0$ is the initial angle of the beam, $R^*$ represents the position value following application of all matrices to the right of the current matrix and the second matrix carrying a factor of $(\pm)\theta_{split}$ adds to the transverse angle for the distribution in the $+\theta_{split}$ direction and subtracts from the $-\theta_{split}$ direction. The final transform, given the symmetry of the setup, taking d to define the cavity length and with $|B-A|=|D-C|$ becomes:

$$\begin{bmatrix} R_1 \\ \theta_1 \end{bmatrix} = \begin{bmatrix} R_0 + \theta_0 \cdot (d + |E-D|) + \theta_{split} \cdot (d + 2 \cdot |E-D|) \\ \theta_0 \pm 2 \cdot \theta_{split} \end{bmatrix} \quad (15)$$

$$\begin{bmatrix} R_1 \\ \theta_1 \end{bmatrix} = \begin{bmatrix} 1 \pm \frac{\theta_{split} \cdot (d + 2 \cdot |E-D|)}{R_0} & d + |E-D| \\ \frac{\pm 2 \cdot \theta_{split}}{R_0} & 1 \end{bmatrix} \begin{bmatrix} R_0 \\ \theta_0 \end{bmatrix}$$

Additional traversals of the cavity can be expressed by including focusing, omitting the matrix representing propagation to the end detector, and propagating the ray back through the cavity:

$$\begin{bmatrix} 1 & (A-B) \\ 0 & 1 \end{bmatrix} \begin{bmatrix} 1 & 0 \\ \frac{(\pm)\theta_{split}}{R^*} & 1 \end{bmatrix} \begin{bmatrix} 1 & (B-C) \\ 0 & 1 \end{bmatrix} \begin{bmatrix} 1 & 0 \\ \frac{\pm \theta_{split}}{R^*} & 1 \end{bmatrix} \begin{bmatrix} 1 & (C-D) \\ 0 & 1 \end{bmatrix} \begin{bmatrix} 1 & 0 \\ \frac{-1}{f} & 1 \end{bmatrix} \begin{bmatrix} R_0 \\ \theta_0 \end{bmatrix} \quad (16)$$

where the factors above are already defined. Note that this transformation is applied n times, once for each reflection from a cavity mirror. The overall transformation of the input vector can be expressed in terms of the final vector in 15:

$$\begin{bmatrix} R_n \\ \theta_n \end{bmatrix} = \begin{bmatrix} [1 - \frac{d}{f} \pm \frac{\theta_{split} \cdot d}{R_1}] & d \\ \frac{-1}{f} \pm \frac{2 \cdot \theta_{split}}{R_1} & 1 \end{bmatrix}^{n-1} \cdot \begin{bmatrix} R_1 \\ \theta_1 \end{bmatrix} \quad (17)$$

Figure 2 shows the losses following 15 traversals of the cavity described in Table 1, where the change in photon density as a function of distance from the beam center is calculated.

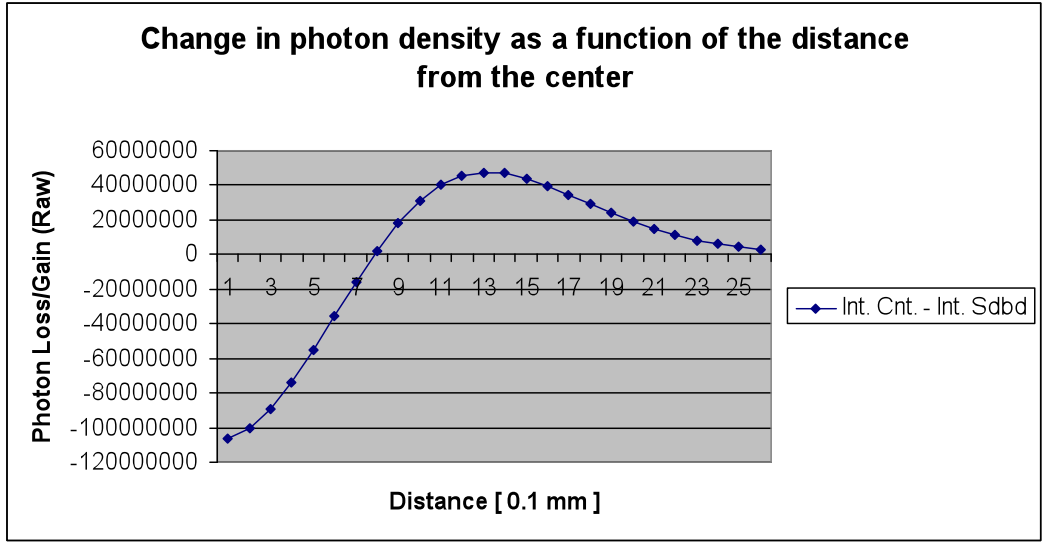

Figure 2: Changes in the photon density after 15 traversals through the field. Each point represents an integral over 0.1 mm, starting from the center, x=0, going out to 3.0 mm.

In this plot, the y axis shows the raw numbers of photons either gained or lost from each of 30 bins having a width of 0.1 mm. The plot starts at the beam center and moves out to a distance of 3.0 mm in one direction. Thus, to extract a total number of photons shifted the absolute value of the distribution must be summed and doubled. The density changes show that photons are lost near the peak (center of the Gaussian) these changes drops to zero around the 0.75 mm mark (the beam waist) and finally shows an increase from the waist out to a position of 3.0 mm. These raw numbers assume a Gaussian function with amplitude of $5 \cdot 10^{+18}$ photons per second.

Figure 3 shows the energy loss from a single central pixel (± 1 μm) as a function of the number of traversals through the cavity.

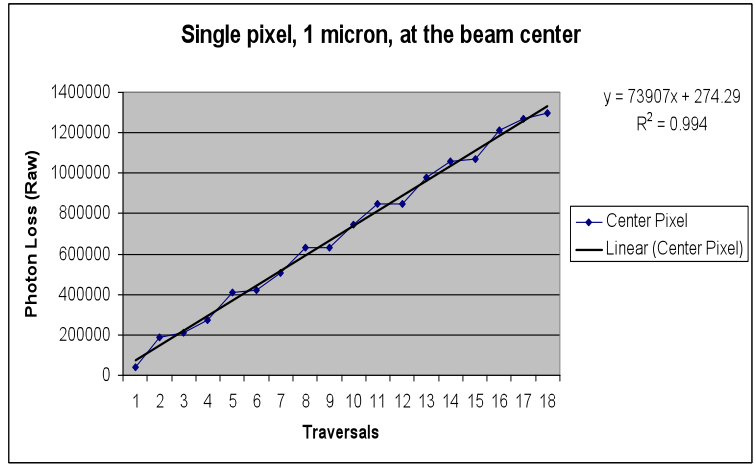

(a)

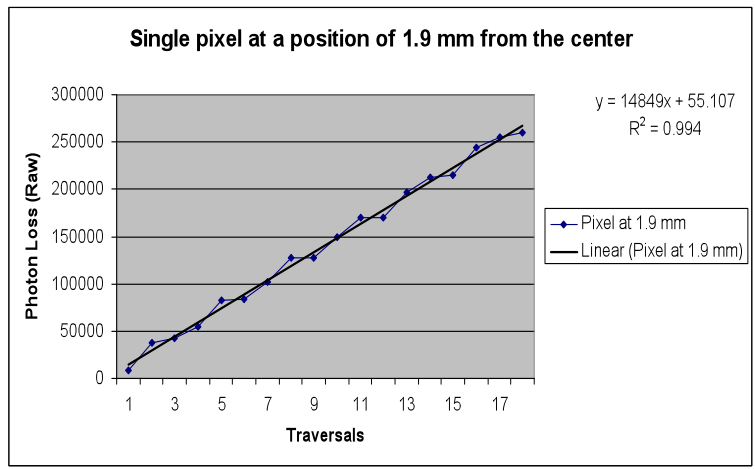

(b)

Figure 3: Plot of the energy loss for a single pixel: (a) at the center of the distribution and (b) in the sideband at a distance of 3.3 mm, as a function of the number of traversals through the field.

An important result is that <u>the loss shows a linear growth with each traversal, even in a confocal cavity</u>. Using the fit, a prediction can be made for the loss after some 12000 traversals:

$For\ the\ central\ pixel:$

$$73907 \frac{photons}{transit} \cdot 12000\ transits + 274 = 8.9E + 08\ photons$$

(18)

$For\ the\ Sideband\ pixel:$

$$14849 \frac{photons}{transit} \cdot 12000\ transits + 55 = 1.8E + 08\ photons$$

(19)

Comparing to the total number of photons in these pixels:

$$For\ the\ center\ P = 5.42E+15$$

$$\frac{8.9E+08}{5.42E+15} = 1.6E-07 \tag{20}$$

$$For\ the\ sideband\ P = 1.92E+14$$

$$\frac{1.8E+08}{1.92E+14} = 9.4E-07$$

The final numbers, $3.9\ 10^{-08}$ and $1.5\ 10^{-07}$, are above the $1/\sqrt{n}$ noise of $1.35\ 10^{-08}$ and $7.2\ 10^{-08}$ respectively. Integrating for a period of time equal to $3.0\ 10^{+04}$, reduces the noise to the level of $(1.6 - 5.2)\ 10^{-10}$. Thus a single pixel is sufficient to probe down to levels of $g_a \approx 2.1 \cdot 10^{-8}$. over an integrated time of about 8 hours.

From Figure 3 the function $f(n)$, recall equation 27, defining the effective separation of tracks from their original trajectory with each traversal through the magnetic field can be extracted. It is straightforward to see that this function for a confocal cavity is $\sim \sqrt{n}$ or linear with $f(n)^2$.

Since the coupling essentially promotes photons to a higher angular momentum state, resulting in additional energy in the non-gaussian modes, photon-axion interactions provides a unique signal. In place of simply looking at the change in the intensity of the light at the center, one could exploit this angular momentum change by looking at the difference between the number of photons in the central region, out to the radius of the beam, and the side band regions, defined as just past the radius going out 2 – 3 times the beam waist. Equation 20 defines a signal based on using the shifting of photons to the sidebands:

$$A \equiv \int_{Center} P_D - \varepsilon$$

$$B \equiv \int_{Sideband} P_D + \varepsilon \tag{21}$$

$$Using\ the\ fact.\ \int_{Center} P_D - \int_{Sideband} P_D \approx \frac{1}{10} \int_{Center} P_D$$

$$A - B \approx [0.9 \cdot \int_{Center} P_D] - 2 \cdot \varepsilon \tag{22}$$

Here the change in density is expressed as $P_D \pm \varepsilon$, where $\varepsilon$ represents the small change due to photon-axion interactions. What is interesting about this signal, unlike a drop in intensity as measured by previous Axion searches, it is not easily mimicked by spurious effects such as diffractive losses or power and pointing instabilities.

Figure 4 shows the difference between the integrated central region, defined from $-0.5 \cdot w_0$ to $+0.5 \cdot w_0$, and the integrated sidebands, defined by $w_0 \pm (1.0\ mm + 3\ w_0)$.

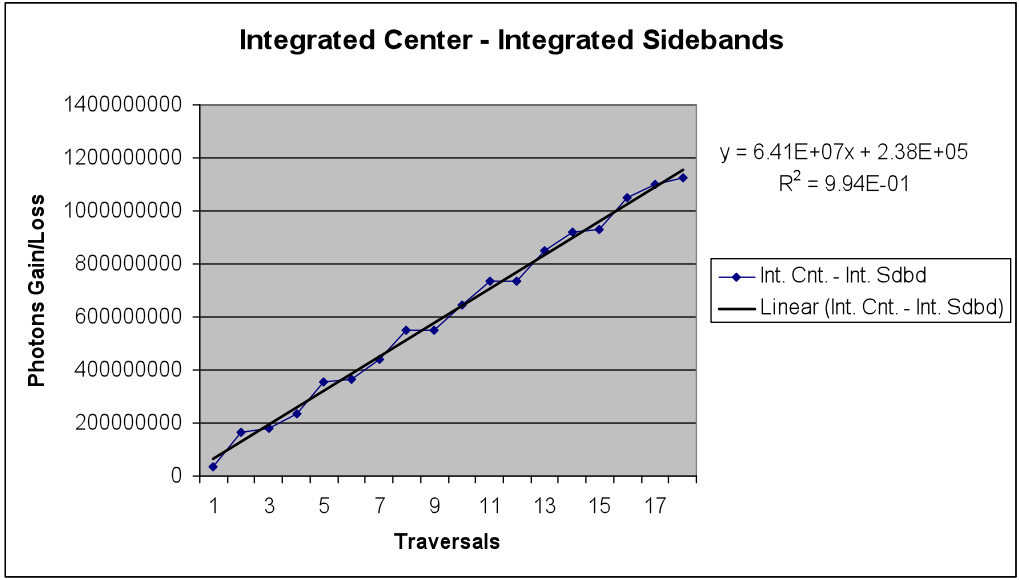

Figure 4: The integrated central region minus the integrated sidebands as a function of the number of traversals.

Using the fit, a prediction can be made for the loss after 12000 traversals:

$$For\ the\ \mathrm{integrated}\ center - the\ \mathrm{integrated}\ sidebands:$$
$$6.41E+07\frac{photons}{transit}\cdot 12000\ transits + 24793 = 7.7E+11\ photons \tag{23}$$

and can be compared to the integrated number of photons ($5 \cdot 10^{18}$) over the beam to yield a fractional change of $1.54 \cdot 10^{-7}$. The noise level comes in at around $4.5 \cdot 10^{-10}$ and implies that a coupling of $g_a \sim 5.4 \cdot 10^{-8}$ GeV$^{-1}$ can be probed within 1 sec of data taking. Integrating for the $3 \cdot 10^4$ sec makes possible measuring couplings down to $g_a \sim 4.1 \cdot 10^{-9}$ GeV$^{-1}$. This of course assumes shot noise limit, but it is <u>promising that with the modest cavity parameters in Table 1 and an integration time of only 8.33 hours such a limit can be achieved.</u>

**Case 2: Planar-Concaved Cavity**

Alternative cavity scenarios have also been simulated to determine the impact of focusing on the defined signal. Recognizing that the final transformation matrix for multiple traversals, see equation 17 above, involves products of $d \cdot (1 - d / f)$ multiplying both the splitting and injection angles, it is easy to anticipate the impact that focusing has on the relative positions of new beams. Ultimately, this relative position is paramount to the size of the effect as seen in equation 8, thus the signal is highly sensitive to focusing.

Stable cavities use focusing to counter natural beam divergence. It should be reiterated that changes in divergence (broadening) is profoundly different from splitting. As seen in Case 1, splitting can grow even when the beam waist is accurately maintained over a very long distance.

Letting the focal length of mirror M2 go to infinity and extracting the light through the concaved, partially reflecting, entrance mirror M1, the extracted signal as a function traversals:

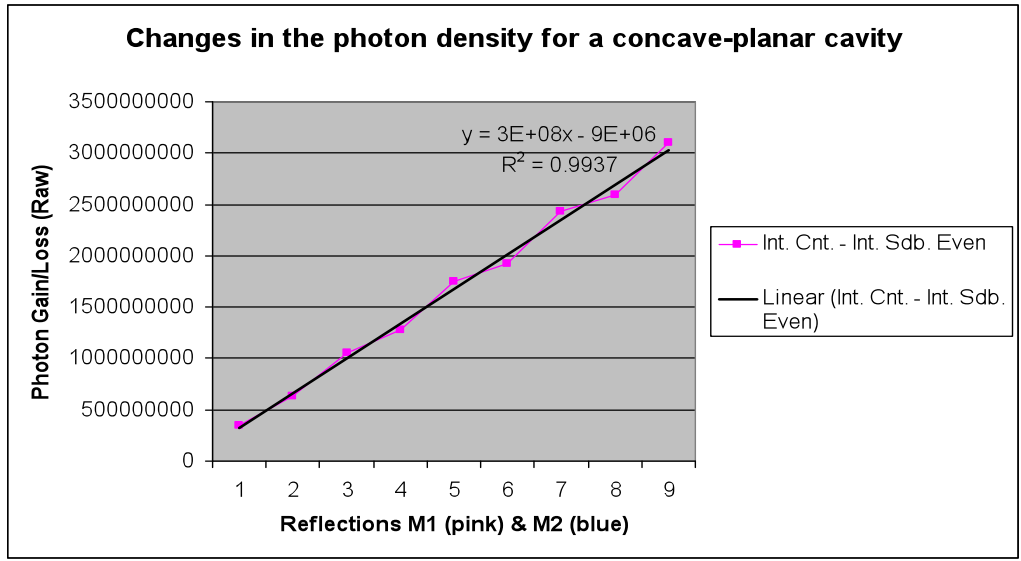

Figure 5: Integrated Center – Integrated Sidebands for the case where the focus of M1 is taken as infinite (concaved-planar cavity scenario).

From the fits a prediction for the change in density following some number of traversals can be predicted. The fit gives a loss about 5 times greater than for the case of a confocal cavity.

**Case 3: Convex-Concaved Cavity**

A more promising alternative involves the use of a concave-convex scenario. In this case, the second mirror (M2) is defocusing.

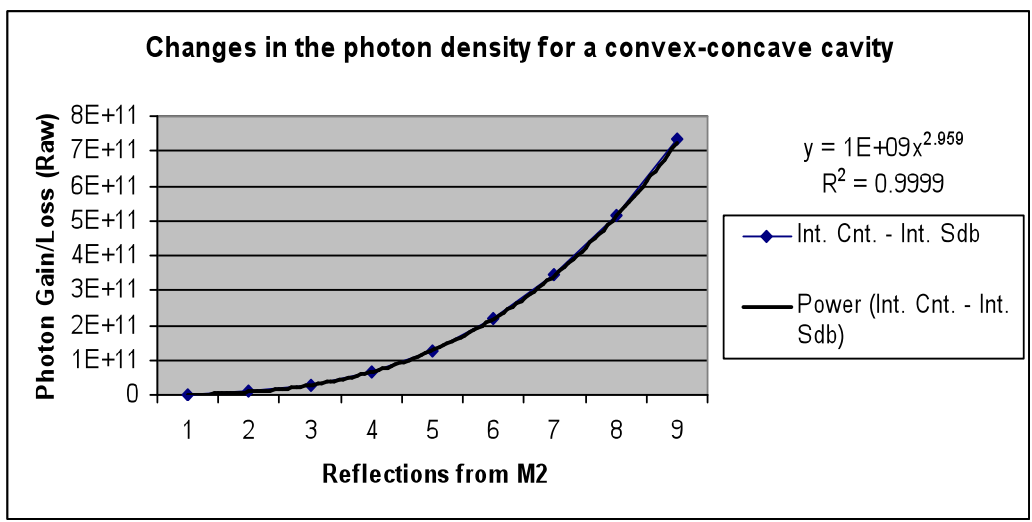

Figure 6: Integrated Center – Integrated Sidebands as a function of traversal through a convex-concave cavity showing the fits to the odd and even traversals separately.

To form a stable cavity, the focus of M2 is simply that of M1 (25.0 m / 2.0) minus half the length of the cavity (14.0 m / 2.0). As in the concave-planar case, the light exits from the partially reflective, concaved mirror thus ever other traversal is actually plotted. What is apparent here is that the function above grows more rapidly than linear.

In this cavity type, the predictions for high numbers of bounces are substantially different and the function $f(n)$ grows as $\sim 2 \cdot \sqrt{2}$. This is the square of the average separation of the chief rays as a function of the number of traversals; seen for the case where the mirrors are both planar. In the convex-concave cavity, the distribution grows quite rapidly. After just 2000 transitions with the fit above, the predicted change in density is:

$$For\ the\ \text{integrated}\ center - the\ \text{integrated}\ sidebands:$$

$$1.0E+09 \cdot 1000^{2.959}\ photons\ = 7.53E+17\ photons \quad (24)$$

If the shot noise limits the background, as before, the measurable coupling strength $g_a$ can be reduced from $10^{-6}$ GeV$^{-1}$ down to about $5.5\ 10^{-11}$ GeV$^{-1}$ with a 1 sec measurement. Thus a time integration of roughly 8.33 hours makes measurable a coupling of $\sim 4.1\ 10^{-12}$; and it becomes possible to contemplate both lower coupling strengths and nonzero masses for the axion in a region not yet measured.

**IV Measurable Mass Range**

The calculations above have assumed something called maximum mixing which implies the axion mass meets the condition described in reference [3]. If the axion has a non-zero mass, the measurable signal is reduced by a factor proportional to the new mixing angle:

$$\tan(2 \cdot \phi) = \frac{2 \cdot Q_M}{Q_\gamma - Q_a}$$

*Where the parameters are defined:*

$$Q_M \approx \omega g_a B^e \approx 1ev \cdot 10^{-21} ev^{-1} \cdot 1T \cdot 195 \frac{ev^2}{T} \approx 10^{-19} ev^2 \quad (25)$$

$$Q_\gamma = \omega^2 \frac{7\alpha}{45\pi} \left(\frac{B^e}{B_{crit}}\right)^2 \approx 3.19 \cdot 10^{-23} ev^2$$

$$Q_a = -m_a^2$$

It is clear that the mixing depends on the photon frequency $\omega$, the strength of the external field $B^e$, and the axion-photon coupling constant $g_a$. For the parameters given in Table 1, the measurable non-zero axion mass is up to $m_a = 5.7\ 10^{-10}$. Equation 25 can also be used to evaluate the mixing for any assumed axion mass given the appropriate input parameters.

Currently, the axion's allowed mass range is up to ~ $10^{-5}$ ev.  Probing a specified mass range becomes a matter of trading off between: the number of traversals achievable, the focal lengths of the mirrors used, the time over which data is collected, the strength and length of the external field, the axion-photon coupling constant and any focusing elements, outside of the cavity but before the detector, that further reduce the beam waist at the detector.

The simulations above prove especially important as a tool for understanding what limits may be achieved with a cavity experiment.  Consider now a cavity constructed around readily available magnets, such as the Brookhaven National Laboratory (BNL) RHIC Quadrupoles.  Table 2 shows some of the parameters for a conceivable experiment.

Table 2:

| Cavity Type | Cavity Length | Magnetic Field Length | ∇B Strength | Laser Wavelength | Laser Energy | Mirror Radius | Number of Transits | $\theta_{split}$ ~ $10^{-10}$ ($g_a = 10^{-10}$) |
|---|---|---|---|---|---|---|---|---|
| Convex-Concaved | 14 m | 1 m | 100 T/m | 1064 nm | 1 W | 25 m | $3.0 \cdot 10^4$ | $2 \cdot 10^{-14}$ |

Where the curvature for the concaved mirror is specified and a stable condition for the curvature of the convex mirror is imposed.  The density changes can be estimated for a single pass system:

$$P_D - (P_D' + P_D'') \approx \frac{5}{6} \cdot 10^{+18} [\frac{2 \cdot 10^{-14} \cdot 14}{7.5 \cdot 10^{-4}}]^2 \approx 1.18 \cdot 10^{-1} \qquad (26)$$

This gives the initial signal level, the integrated change in the photon density near the center of the beam.  As before, light exits through M1 and thus the number of transits must be divided by two.  Using Table 2 to calculate the integrated center minus the integrated sidebands:

$$For\ the\ \mathrm{int}egrated\ center - the\ \mathrm{int}egrated\ sidebands:$$
$$2.2E+00 \cdot 15000^{2.959}\ photons = 5.00E+12\ photons \qquad (27)$$

Note: the coupling constant here was taken to be $10^{-10}$ GeV$^{-1}$, the CAST limit.  With a noise level of ~ 4.5 $10^{-10}$ photons, the signal is sufficient to measure such a coupling within a single second of data taking.  Integrating over time, 35 days or 3 $10^6$ seconds, gives a measurable coupling constant at a level of 5.1 $10^{-14}$ GeV$^{-1}$.

A trade off can be made between going to lower couplings and probing higher mass values. Further exploration of this topic is left for a future publication.

## V. Conclusion

The extension of the theory behind photon-axion coupling in an external magnetic field to terrestrial experiments holds much promise for improving the sensitivity of axion searches. The splitting effect that is predicted can be enhanced in such a cavity environment due to something known as bifurcation. It has been shown, through careful simulations, that a measurable effect can grow as the number of traversals of the field to the third power. With current technology, a robust experiment could probe scales previous unavailable and stand a reasonable chance of hitting the theoretical coupling constants.

## VI. Acknowledgements

A special thanks is given to Mikhail Khankhasayev who assisted through reading and editing this manuscript. This work was made possible by a grant from the National Science Foundation and support from the Florida A&M University Department of Physics.

## V. References:

[1.] G. Rafetti and L. Stodolsky, *Mixing of the Photon with Low-mass Particles*. Phys. Rev. D 37 (1988) 1237.

[2.] E. I. Guendelman, *Photon and Axion splitting in an inhomogeneous Magnetic field*, Phys. Lett. B662, 445-448 (2008)

[3.] Mikhail Khankhasayev and Carol Scarlett, *On the Possibility to Observe Photon-Axion Mixing Effects in a Cavity Experiment*, ArXiv 1201.6008v2 (2012)

[4.] Y. Semertzidis, *Coherent Production of Light Pseudoscalars (Axions) Inside a Magnetic Field with a Polarized Laser Beam*, University of Rochester, (1989)

[5.] R. Cameron, *Search for New Photon Couplings in a Magnetic Field*, University of Rochester, (1992)

[6.] U. Gastaldi *et al.*, *PVLAS Results*, INFL-LNL (REP) 206/05 (2005).

[7.] Battesti *et al.*, *The BMV experiment: a novel apparatus to study the propagation of light in a transverse magnetic field*, Eur. Phys. J. D 46, 323-333 (2008)

**APPENDIX:**

To better understand the difference between the predictions for a bifurcation effect and those of a Pascal Triangle, consider the simplified example below. In this case, the propagation through the field is treated linearly, no adjustment for curvature, the mirrors are taken to be planar and the field extends over the entire region of the cavity (not realistic but illustrative).

If the Pascal triangle is to describe the position of new beams emerging through photon-axion interactions, it can be compared to position information extracted by applying a bifurcation.

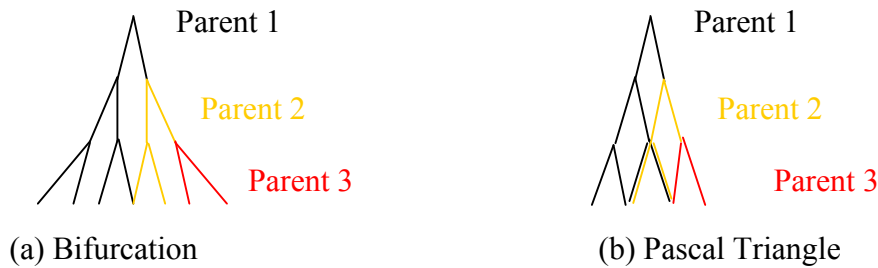

Figure 8: Comparison of a bifurcation of a beam, following three traversals of a cavity, to an estimate using a Pascal Triangle.

In both scenarios, a single beam splits into two beams one with positive momentum in the transverse direction ( + 1 $P_y$) the other with negative momentum ( − 1 $P_y$). Following a reflection, the bifurcation diagram shows two beams in (a) emerging with the same transverse momentum acquired through interacting with axions and the external magnetic field. Reentering the field, these two beams split into four new beams emerging with transverse momentum $+ 2 P_y$, $0 P_y$, $0 P_y$ and $- 2 P_y$. For the Pascal triangle, following the reflection, the two beams reenter with transverse momentum of $0 P_y$. These beams split into four beams having transverse momentum of $+ 1P_y$, $-1 P_y$, $+1 P_y$ and $- 1P_y$.

For only a few traversals of the field, such differences may seem insignificant. However, if transverse momentum were not conserved at the mirror the shifting of energy away from the center would grow as the square-root of the distance traveled. This is significantly less than what simulations, assuming conservation of transverse momentum following reflection, predict for even the confocal cavity; which shows linear growth (after 10000 m the signal according to the Pascal triangle has increased by only 100).